# Uncertainty-based pressure field reconstruction from PIV/PTV flow measurements with generalized least-squares

Jiacheng Zhang[1], Sayantan Bhattacharya[1], and Pavlos P. Vlachos[1,2]

[1]School of Mechanical Engineering, Purdue University, West Lafayette, IN 47907, USA

[2]Weldon School of Biomedical Engineering, Purdue University, West Lafayette, IN 47907, USA

pvlachos@purdue.edu

## Abstract

A novel uncertainty-based pressure reconstruction method is proposed to evaluate the instantaneous pressure fields from velocity fields measured using particle image velocimetry (PIV) or particle tracking velocimetry (PTV). First, the pressure gradient fields are calculated from velocity fields, while the local and instantaneous pressure gradient uncertainty is estimated from the velocity uncertainty using a linear-transformation based algorithm. The pressure field is then reconstructed by solving an overdetermined linear system which involves the pressure gradients and boundary conditions. This linear system is solved with generalized least-squares (GLS) which incorporates the previously estimated variances and covariances of the pressure gradient errors as inverse weights to optimize the reconstructed pressure field. The method was validated with synthetic velocity fields of a 2D pulsatile flow and the results show significantly improved pressure accuracy with an error reduction of as much as 250% compared to the existing baseline method of solving the pressure Poisson equation (PPE). The GLS was more robust to the velocity errors and provides greater improvement with spatially correlated velocity errors. For experimental validation, the volumetric pressure fields were evaluated from a laminar pipe flow velocity field measured using 3D PTV. The GLS reduced the median absolute pressure errors by as much as 96%.

Keywords: generalized least-squares, instantaneous pressure reconstruction, uncertainty estimation

# Nomenclature

| | | | |
|---|---|---|---|
| x, y, z | spatial coordinates | $p_{ref}$ | reference pressure |
| $\Delta x, \Delta y$ | grid sizes | $\epsilon_{p_{ref}}$ | error in the reference pressure |
| $\Delta t$ | time interval between consecutive frames | $\boldsymbol{p_{grad,u}}$ | pressure gradient calculated from the measured velocity |
| u, v, w | velocity components | $\boldsymbol{p_{grad,t}}$ | true pressure gradient |
| $\boldsymbol{u}$ | velocity vector | $\boldsymbol{\epsilon_{\nabla p}}$ | error in the calculated pressure gradient |
| $\boldsymbol{u_m}$ | measured velocity | $\boldsymbol{\sigma_{\nabla p}}$ | standard deviation of pressure gradient errors |
| $\boldsymbol{u_t}$ | true velocity | $\Sigma_{\nabla p}$ | covariance matrix of pressure gradient errors |
| $\boldsymbol{\epsilon_u}$ | error in measured velocity | $\epsilon_{pgradx}$ | streamwise pressure gradient errors |
| $\Sigma_u$ | covariance matrix of velocity errors | $\sigma_{pgradx}$ | standard deviation of streamwise pressure gradient errors |
| $\boldsymbol{\sigma_u}$ | standard deviation of velocity errors | $\rho_{pgradx}$ | auto-correlation coefficients between streamwise pressure gradient errors |
| $\rho_{u1,u2}$ | auto-correlation coefficient between velocity errors | $\rho$ | fluid density |
| $Cov_{u1,u2}$ | covariance between velocity errors | $\mu$ | fluid dynamic viscosity |
| $p$ | pressure | STD | standard deviation |
| $\epsilon_p$ | error in the reconstructed pressure | RMS | root mean square |

# 1 Introduction

Measurement of pressure in a fluid flow is important in engineering applications as well as in investigations of flow physics. Pressure measurement devices such as wall pressure ports, static tubes, pressure-sensitive painting (PSP) etc, are either invasive, provide point measurements or a surface distribution (McKeon and Engler, 2007). Further, most pressure measurement techniques have limitations in dynamic range and resolvable frequency bandwidth. With the development of flow measurement techniques such as particle image velocimetry (PIV) and particle tracking velocimetry (PTV), the velocity fields can be obtained and utilized for instantaneous pressure evaluation (Fujisawa et al. 2005; Liu and Katz 2006; Charonko et al. 2010; Neeteson and Rival 2015; Huhn et al. 2016). Most pressure reconstruction methods require two major steps to calculate the pressure fields from velocity measurements. The pressure gradient fields are first evaluated from the velocity fields using the Navier-Stokes momentum equation, which are then spatially integrated to reconstruct the pressure fields. For incompressible flows, the momentum equation can be expressed in the following form:

$$\nabla p = -\rho \frac{D\boldsymbol{u}}{Dt} + \mu \nabla^2 \boldsymbol{u}, \qquad (1)$$

where $p$ is the pressure, $\boldsymbol{u}$ is the velocity, $\rho$ and $\mu$ are the density and dynamic viscosity of the fluid, respectively. $\nabla$ represents the divergence operator and $\nabla^2$ is the Laplacian operator. $\frac{D\boldsymbol{u}}{Dt}$ is the material acceleration which can be evaluated using the Eulerian approach from gridded velocity data (Fujisawa et al. 2005; de Kat et al. 2009; Charonko et al. 2010; Tronchin et al. 2015) or the Lagrangian approach from particle tracks (Neeteson and Rival 2015; Gesemann et al. 2016; Huhn et al. 2016). For pressure integration, one common approach is path-integration (also referred to as spatial-marching) which integrates the pressure gradient along paths across the flow domain (Liu and Katz 2006; Dabiri et al. 2014; Tronchin et al. 2015). Most path-integration schemes employ redundant number of paths to reduce the influence of erroneous pressure gradient values. The path-integration approach is rarely employed for 3D flow data due to its high computational cost. The most widely used pressure integration approach is solving the pressure Poisson equation (PPE) (Fujisawa et al. 2005; de Kat et al. 2009; Violato et al. 2011; Neeteson and Rival 2015; Schneiders and Scarano 2016). The performances of PIV-based pressure calculation methods have been explored by Charonko et al. (2010), and a comparative assessment of pressure field reconstructions from PIV and PTV have been performed by van Gent et al. (2017) based on a simulated experiment.

However, the pressure reconstruction from flow measurements have inherent limitations. First, spatio-temporally resolved velocity measurements are normally required to accurately reconstruct the instantaneous pressure fields. Schneiders et al. (2014) proposed a vortex-in-cell based algorithm for time-supersampling of 3D PIV measurements. Schneiders and Scarano (2016) developed a dense velocity reconstruction method for tomographic PTV. Gesemann et al. (2016) employed a B-splines based global minimization method to obtain high-resolution pressure fields from particle tracks by STB method. Moreover, the accuracy of the reconstructed pressure fields is significantly affected by the measurement error in the velocity fields (Charonko et al. 2010; Azijli et al. 2016; Pan et al. 2016). The velocity errors propagate through the pressure gradient calculation and pressure integration to the reconstructed pressure fields with an error amplification of as much as 100 times depending on the type of flow, the governing equation, and the prescribed boundary conditions (Charonko et al. 2010). Smoothing and filtering can be employed to mitigate errors during pressure gradient evaluation (Charonko et al. 2010; Wang et al. 2016; Schiavazzi et al. 2017). In order to reduce error propagation during pressure integration, Tronchin et al. (2015) and Jeon et al. (2018) employed approaches which divided the flow field into subdomains with respect to local velocity reliabilities, then performed pressure integration in the subdomains in descending order of the reliabilities. Consequently, the evaluated pressure in a more reliable subdomain was not affected by the erroneous velocity measurements in a less reliable subdomain. This type of approaches is particularly effective for flow fields that can be segmented into regions with different levels of measurement accuracy. One example is the flow field around an airfoil which can be divided into outer-region, wake-region, near-body region, and near-edge region with descending accuracy (Jeon et al. 2018). In these works, the reliability of each subdomain was defined by the Frobenius norm of the velocity gradient or pressure gradient tensor.

Since the pressure reconstruction process is significantly affected by the errors in the estimated velocity field, the uncertainty bounds on each velocity measurement can also be used as a measure of reliability to subsequently optimize the error propagation in pressure field estimation. The standard uncertainty is estimated as the standard deviation of the error distribution and provides a bound on the error distribution with certain confidence. For flow measurements using PIV, recent developments have enabled the uncertainty estimation for each velocity vector in the flow field (Charonko et al. 2010; Xue et al. 2014, 2015; Bhattacharya et al. 2017; Schneiders and Sciacchitano 2017; Bhattacharya et al. 2018). The local and instantaneous velocity uncertainty has been used to denoise the velocity fields using a spatial averaging scheme (Wieneke 2017) and can be propagated to estimate the uncertainty in the calculated pressure gradient fields as well as the reconstructed pressure fields (Azijli et al. 2016). However, to the best of the authors' knowledge, the uncertainty information has not been utilized to improve the pressure integration.

In the present study, a robust pressure reconstruction method is proposed which employs the velocity uncertainty to improve the accuracy of reconstructed pressure fields. A linear-transformation based uncertainty propagation algorithm is developed to acquire the pressure gradient uncertainty as variances and covariances. The pressure integration on a discretized domain is formulated as an overdetermined linear system involving pressure gradients and pressure boundary conditions. The generalized least-squares (GLS), which is the best unbiased linear estimator (BLUE) (Aitken 1935), is employed to reconstruct the pressure fields with the estimated pressure gradient uncertainty. The performance of the GLS pressure reconstruction method is tested with synthetic velocity fields and applied to volumetric flow data measured using PTV.

## 2 Methodology

### 2.1 Uncertainty-based pressure reconstruction with generalized least-squares

The procedure of uncertainty-based pressure reconstruction with GLS is presented in Fig. 1. The pressure gradient calculation is described in Sect. 2.1.1. The linear-transformation based uncertainty propagation algorithm is introduced in Sect. 2.1.2. The pressure integration methods and numerical schemes are introduced in Sects. 2.1.3 and 2.1.4, respectively. In Sect. 2.1.5, a velocity-divergence based uncertainty estimation algorithm is introduced as a substitute to provide velocity uncertainty for GLS reconstruction.

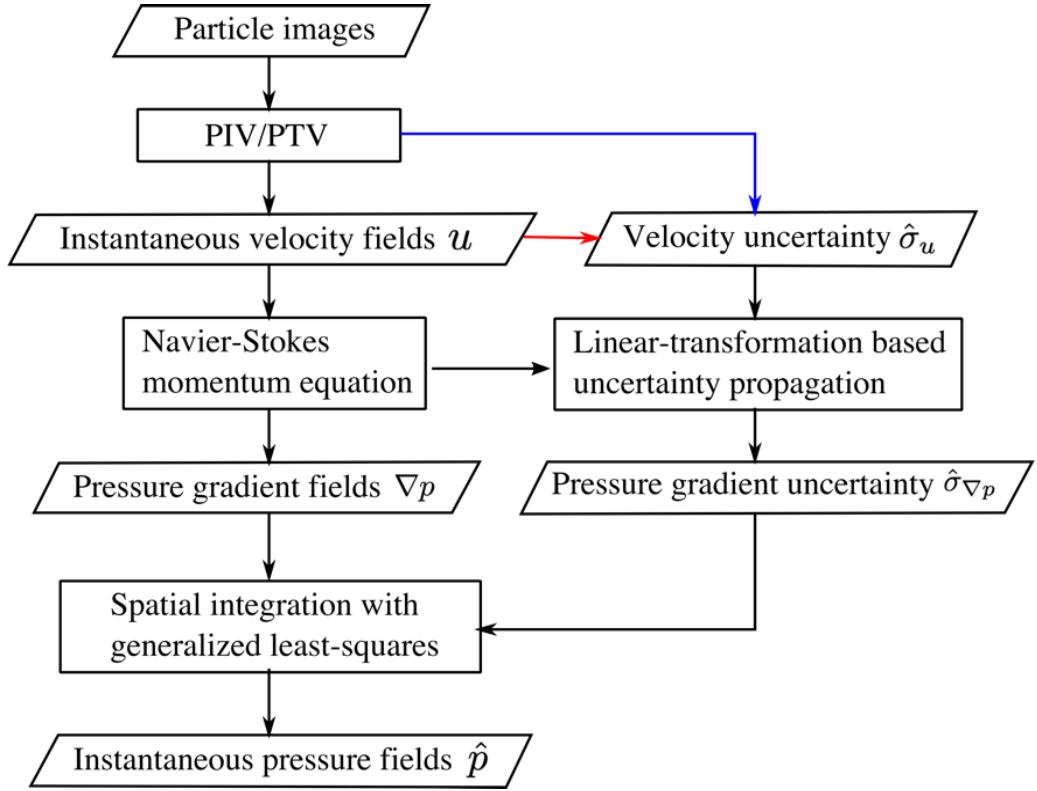

**Fig. 1 Procedure of the uncertainty-based GLS pressure reconstruction. The velocity uncertainty can be obtained from the flow measurements (blue arrow) or estimated from the velocity fields (red arrow) using the divergence-based algorithm.**

*2.1.1 Pressure gradient calculation*

The pressure gradient field is estimated by substituting the velocity field in the Navier-Stokes equations. As the velocity fields employed in the present study were on Cartesian grids, the pressure gradient fields were calculated using the Eulerian approach as:

$$\boldsymbol{p}_{grad,u} = -\rho \left( \frac{\partial \boldsymbol{u}_m}{\partial t} + \boldsymbol{u}_m \cdot \nabla \boldsymbol{u}_m \right) + \mu \nabla^2 \boldsymbol{u}_m, \tag{2}$$

where $\boldsymbol{p}_{grad,u}$ is the pressure gradient evaluated from the measured velocity $\boldsymbol{u}_m$, $\frac{\partial \boldsymbol{u}_m}{\partial t}$ is the local acceleration, and $\boldsymbol{u}_m \cdot \nabla \boldsymbol{u}_m$ is the convective acceleration. The spatial and temporal derivatives were calculated using the second-order central difference scheme for grid points in the domain. The first-order one-sided scheme was used at the boundaries for spatial derivatives and at the first and last frames for the temporal derivatives. The evaluated pressure gradient values were on the same grid points as the velocity. The pressure gradient calculation is carried out using matrix/vector operations. At each frame, the velocity field and the calculated pressure gradient field were organized as column vectors containing all the components from all the grid points, i.e., $\boldsymbol{u_m} = [u\ v\ w]^T$ and $\boldsymbol{p_{grad,u}} = \left[\frac{dp}{dx}\ \frac{dp}{dy}\ \frac{dp}{dz}\right]^T$ for volumetric flow data. The discretized gradient and Laplacian operators were 2D matrices with respect to the selected difference schemes. The following equation demonstrates the calculation of pressure gradient for the i$^{th}$ frame as:

$$\boldsymbol{p_{grad,u}^i} = -\rho\left(\frac{u_m^{i+1} - u_m^{i-1}}{2\Delta t} + \boldsymbol{u_m^i} \odot (\nabla \boldsymbol{u_m^i})\right) + \mu \nabla^2 \boldsymbol{u_m^i} \tag{3}$$

where the superscript denotes the frame number, $\Delta t$ is the time difference between consecutive frames, and $\odot$ represents the Hadamard product (entry-wise product).

### 2.1.2 Pressure gradient uncertainty estimation using linear transformations

The proposed pressure reconstruction methods requires both pressure gradients and the pressure gradient uncertainty for pressure integration. To estimate the uncertainty in $\boldsymbol{p_{grad,u}}$, we implemented a linear-transformation based uncertainty propagation algorithm.

Considering the measurement errors, $\boldsymbol{u_m}$ and $\boldsymbol{p_{grad,u}}$ can be decomposed into true components and error components as,

$$\boldsymbol{u_m} = \boldsymbol{u_t} + \boldsymbol{\epsilon_u}, \tag{4}$$

$$\boldsymbol{p_{grad,u}} = \boldsymbol{p_{grad,t}} + \boldsymbol{\epsilon_{\nabla p}}, \tag{5}$$

where the subscript t suggests the true component, $\boldsymbol{\epsilon_u}$ and $\boldsymbol{\epsilon_{\nabla p}}$ are error components of velocity and pressure gradient, respectively. With the assumption that the velocity fields are acquired with sufficient spatiotemporal resolutions such that the numerical truncation errors are negligible, the true pressure gradient can be evaluated as,

$$\boldsymbol{p_{grad,t}} = -\rho\left(\frac{\partial \boldsymbol{u_t}}{\partial t} + \boldsymbol{u_t} \cdot \nabla \boldsymbol{u_t}\right) + \mu \nabla^2 \boldsymbol{u_t}. \tag{6}$$

Combining Eqs. 2, 4 and 6, the pressure gradient error can be obtained in terms of the measured velocity $\boldsymbol{u_m}$ and the velocity error $\boldsymbol{\epsilon_u}$ as follows:

$$\boldsymbol{\epsilon_{\nabla p}} = -\rho\left(\frac{\partial \boldsymbol{\epsilon_u}}{\partial t} + \boldsymbol{u_m} \cdot \nabla \boldsymbol{\epsilon_u} + \boldsymbol{\epsilon_u} \cdot \nabla \boldsymbol{u_m} - \boldsymbol{\epsilon_u} \cdot \nabla \boldsymbol{\epsilon_u}\right) + \mu \nabla^2 \boldsymbol{\epsilon_u}, \tag{7}$$

Assuming that $\boldsymbol{\epsilon_u}$ is sufficiently less than the velocity $\boldsymbol{u_m}$, the term $\boldsymbol{\epsilon_u} \cdot \nabla \boldsymbol{\epsilon_u}$ can be neglected (Pan et al. 2016). Therefore, equation 7 can be simplified and expressed using matrix/vector operations as:

$$\epsilon_{\nabla p} = M_+ \epsilon_u^{i+1} + M_- \epsilon_u^{i-1} + M \epsilon_u^i,$$

with $M_+ = -\frac{\rho}{\Delta t} I$, $M_- = \frac{\rho}{\Delta t} I$, and $M = -\rho \left( diag(\boldsymbol{u}_m^i) \cdot \nabla + diag(\nabla \boldsymbol{u}_m^i) \right) + \mu \nabla^2$, (8)

where $\epsilon_u^i$ is the column vector containing velocity errors from the i$^{th}$ frame, $I$ is the identity matrix whose dimension is same with the length of $\epsilon_u^i$, and the function $diag()$ constructs a diagonal matrix with the diagonal elements from the given column vector. Equation 8 estimates the pressure gradient errors at the i$^{th}$ frame as the linear transformations of velocity errors at frames i, i-1, and i+1. The transformation coefficients in $M_+$, $M_-$, and $M$ are decided by the measured velocity field and the selected discretization schemes. With the assumption that the velocity errors at different frames are independent, the uncertainty of pressure gradient can be determined as:

$$\Sigma_{\nabla p}^i = M_+ \Sigma_u^{i+1} M_+^T + M_- \Sigma_u^{i-1} M_-^T + M \Sigma_u^i M^T, \tag{9}$$

where $\Sigma_u^i$ and $\Sigma_{\nabla p}^i$ are the covariance matrices of velocity errors and pressure gradient errors at the i$^{th}$ frame, respectively. The $\Sigma_u$ can be constructed based on the velocity uncertainty $\sigma_u$ as $\Sigma_u = diag(\sigma_u^2)$ assuming spatially uncorrelated velocity errors. For correlated velocity errors, the covariances are stored as the off-diagonal elements of $\Sigma_u$.

### 2.1.3 Pressure integration

The pressure field is inherently related to the pressure gradient field as,

$$G\boldsymbol{p} = M_L \boldsymbol{p}_{grad,u}, \tag{10}$$

where $\boldsymbol{p}$ is a column vector of the estimated pressure field, G is the discretized gradient operator constructed with a staggered grid arrangement similar as employed by Jeon et al. (2018), and $M_L$ is the transformation matrix that linearly interpolates $\boldsymbol{p}_{grad,u}$ from grid points to staggered nodes. As demonstrated in Fig. 2, the grid points are represented with circles, the arrow heads indicate the staggered nodes for the interpolated pressure gradients, and the arrow directions indicate the component of pressure gradients. As an example, the staggered grid interpolations in x and y directions at points $x + \Delta x/2, y$ and $x, y + \Delta y/2$ can be written as, $\left.\frac{dp}{dx}\right|_{x+\Delta x/2,y} = \frac{1}{2}\left(\left.\frac{dp}{dx}\right|_{x,y} + \left.\frac{dp}{dx}\right|_{x+\Delta x,y}\right)$ and $\left.\frac{dp}{dy}\right|_{x,y+\Delta y/2} = \frac{1}{2}\left(\left.\frac{dp}{dy}\right|_{x,y} + \left.\frac{dp}{dy}\right|_{x,y+\Delta y}\right)$, respectively. The filled circles (mentioned as reference points in the present study) are the grid points with prescribed pressure values. The following equation relates the reference points with the reconstructed pressure field as

$$L\boldsymbol{p} = \boldsymbol{p}_{ref}, \tag{11}$$

where $L$ is a labeling matrix consisting of 0s and 1s, and $\boldsymbol{p}_{ref}$ is the column vector containing the reference pressure values. With the errors in the pressure gradient fields and the pressure reference values, a linear system can be constructed by combining Eqs. 10 and 11 as

$$\begin{bmatrix} M_L \boldsymbol{p}_{grad,u} \\ \boldsymbol{p}_{ref} \end{bmatrix} = \begin{bmatrix} G & 0 \\ 0 & L \end{bmatrix} \boldsymbol{p} + \begin{bmatrix} M_L \boldsymbol{\epsilon}_{\nabla p} \\ \boldsymbol{\epsilon}_{p_{ref}} \end{bmatrix}, \tag{12}$$

where $\boldsymbol{\epsilon}_{p_{ref}}$ is the column vector of possible errors in the reference pressure values which can be obtained by direct pressure measurement. For the sake of simplicity, we denote $[M_L \boldsymbol{p}_{grad,u} \quad \boldsymbol{p}_{ref}]^T$ by $\boldsymbol{b}$, $\begin{bmatrix} G & 0 \\ 0 & L \end{bmatrix}$ by A, and $[M_L \boldsymbol{\epsilon}_{\nabla p} \quad \boldsymbol{\epsilon}_{p_{ref}}]^T$ by $\boldsymbol{\epsilon}_b$. Since $\boldsymbol{\epsilon}_{\nabla p}$ are uncorrelated with $\boldsymbol{\epsilon}_{p_{ref}}$, the covariance matrix $\Sigma_b$ of the error term $\boldsymbol{\epsilon}_b$ can be obtained as

$$\Sigma_b = \begin{bmatrix} M_L \Sigma_{\nabla p} M_L^T & 0 \\ 0 & \Sigma_{p_{ref}} \end{bmatrix}, \tag{13}$$

where $\Sigma_{p_{ref}}$ is the covariance matrix of $\boldsymbol{\epsilon}_{p_{ref}}$. The pressure field can be estimated ($\widehat{\boldsymbol{p}}$) from Eqn. 12 using GLS which minimizes the following equation as

$$\widehat{\boldsymbol{p}} = \underset{p}{\operatorname{argmin}}(\boldsymbol{b} - A\boldsymbol{p})^T \Sigma_b^{-1}(\boldsymbol{b} - A\boldsymbol{p}). \tag{14}$$

$\widehat{\boldsymbol{p}}$ can be obtained as the solution of the following equation

$$(A^T \Sigma_b^{-1} A) \widehat{\boldsymbol{p}} = A^T \Sigma_b^{-1} \boldsymbol{b}. \tag{15}$$

To avoid singularity, at least one pressure reference point is required. In addition, the variance terms in the diagonal elements in $\Sigma_b$ must be greater than 0. For $\boldsymbol{p}_{ref}$ with negligible uncertainty, the variance of $\boldsymbol{\epsilon}_{p_{ref}}$ can be set at a small fraction of the pressure scale, e.g. $10^{-3}p_0$, where $p_0$ is the characteristic pressure of the flow.

Two other variants of the least-squares reconstruction were also implemented, namely ordinary least-squares (OLS) and weighted least-squares (WLS). Both OLS and WLS calculate pressure fields using Eqn. 15 with extra assumptions compared to GLS. OLS assumes independent and homoscedastic errors in $\epsilon_b$ such that $\Sigma_b$ can be ignored or treated as an identity matrix, while WLS assumes independent errors in $\epsilon_b$, thus the off-diagonal elements (covariances) are zeros in $\Sigma_b$. The least-squares method proposed by Jeon et al. (2018) can be considered as an OLS approach.

The pressure reconstruction by solving the PPE was employed as the baseline method. The following form of PPE was selected which generates the source term as the divergence of pressure gradients as

$$\nabla^2 \boldsymbol{p} = \nabla \cdot \boldsymbol{p}_{grad,u}. \tag{16}$$

where the Laplacian operator $\nabla^2$ was discretized using the second-order central difference scheme with a five-point stencil for planar data and a seven-point stencil for volumetric data. The boundary conditions for solving the PPE were assigned with reference pressure values (Dirichlet BC) or pressure gradients (Neumann BC). For Neumann BC, the pressure gradients were given as the $\boldsymbol{p}_{grad,u}$ calculated at the boundary points.

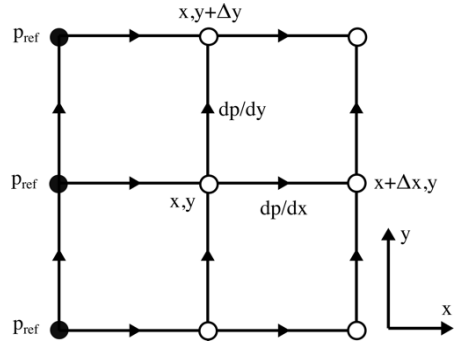

**Fig. 2 Demonstration of the grid arrangement for the pressure integration with generalized least-squares. dp/dx and dp/dy represent the pressure gradients along x and y directions, respectively. $\Delta x$ and $\Delta y$ are the grid sizes. $p_{ref}$ represents the reference pressure values.**

*2.1.4 Numerical schemes for solving linear systems*

For pressure integration using GLS, solving the linear system of Eq. 15 is prohibited as it requires matrix operations involving $\Sigma_b^{-1}$. Although the covariance matrix $\Sigma_b$ is sparse due to the fact that the pressure gradient errors are only correlated within a small neighborhood, the inverse matrix $\Sigma_b^{-1}$ is normally dense with a dimension of approximately $2N_{pts} \times 2N_{pts}$ for planar data or $3N_{pts} \times 3N_{pts}$ for volumetric data, where $N_{pts}$ is the total number of grid points in the flow field. By introducing the vector $\boldsymbol{\lambda} = \Sigma_b^{-1}(\boldsymbol{b} - A\boldsymbol{p})$, the following equation can be solved for $\boldsymbol{p}$ (Rao 1973)

$$\begin{bmatrix} \Sigma_b & A \\ A^T & 0 \end{bmatrix} \begin{bmatrix} \boldsymbol{\lambda} \\ \boldsymbol{p} \end{bmatrix} = \begin{bmatrix} \boldsymbol{b} \\ 0 \end{bmatrix}. \tag{17}$$

which avoids the operations involving the large dense matrix $\Sigma_b^{-1}$. Therefore, more efficient algorithms can be employed to solve the sparse linear system. For OLS and WLS, the pressure fields can be solved from Eq. 15 since $\Sigma_b$ is ignored or treated as a diagonal matrix such that $\Sigma_b^{-1}$ is also a diagonal matrix.

To solve the linear systems of Eqs. 15-17, different numerical schemes were selected depending on the size of the flow data. For planar data and small volumetric data, SuperLU, a general-purpose library for the direct solution of large, sparse, nonsymmetric systems of linear equations (Li 2005), was employed. For large volumetric flow data, the linear systems were solved using Conjugate Gradient iteration (Björck 1996).

### 2.1.5 Divergence-based velocity uncertainty estimation

For some flow measurement techniques such as volumetric PIV and PTV, the estimation of local and instantaneous velocity uncertainty is still unestablished or limited. To inform the uncertainty-based GLS pressure reconstruction, a divergence-based approach can be employed to estimate the velocity uncertainty directly from the velocity fields. For incompressible flow, the velocity errors cause nonzero velocity divergence as:

$$\nabla \cdot \boldsymbol{\epsilon}_u = \nabla \cdot \boldsymbol{u}_m, \tag{18}$$

which can be solved in a least-squares sense to estimate the velocity errors $\hat{\boldsymbol{\epsilon}}_u$ as:

$$\hat{\boldsymbol{\epsilon}}_u = (\nabla \cdot)^T (\nabla^2)^{-1} (\nabla \cdot \boldsymbol{u}_m) \tag{19}$$

This approach was employed by (Zhang et al. 2019) to estimate velocity errors which were then propagated for $\epsilon_{pgrad}$ to inform the weighted least-squares pressure reconstruction. In the present study, the velocity uncertainty at each grid point of each frame can be estimated as the weighted standard deviation (WSTD) of $\hat{\boldsymbol{\epsilon}}_u$ from the spatiotemporally neighboring

points as:

$$\hat{\sigma}_u = \sqrt{\frac{\sum w(\epsilon_u)^2}{\sum w}} \text{ with}$$

$$w = \exp\left(-\frac{1}{2}\left(\frac{\Delta s_r}{\Delta x}\right)^2 - \frac{1}{2}\left(\frac{\Delta s_t}{\Delta t}\right)^2\right), \tag{20}$$

where $\Delta s_r$ and $\Delta s_t$ are the spatial and temporal separations from the neighboring points to the point of interest, respectively. Only the points within the $\Delta s_r \leq \Delta x$ and $\Delta s_t \leq \Delta t$ neighborhood are employed for the $\hat{\sigma}_u$ calculation to ensure the local and instantaneous dependency of $\hat{\sigma}_u$.

### 2.2 Synthetic flow fields

The synthetic velocity fields of a 2D pulsatile flow between two infinite parallel plates were employed to test the uncertainty estimation algorithm and assess the performances of the pressure reconstruction methods. The pulsatile flow is driven by the oscillating streamwise pressure gradient as

$$\frac{dp}{dx} = \rho K + \gamma \rho K \cos\omega t, \tag{21}$$

and the streamwise velocity profile can be expressed as

$$u = u_{max}\left(1 - \frac{y^2}{h^2}\right) + \frac{\gamma K}{i\omega}\left(1 - \frac{\cosh(y/h\sqrt{i}\lambda)}{\cosh(\sqrt{i}\lambda)}\right)\exp(i\omega t) \tag{22}$$

with $\lambda = h\sqrt{\frac{\rho}{\mu\omega}}$, where $\gamma$ is the ratio between the magnitude of the steady pressure gradient component and the amplitude of the oscillating pressure gradient component, K is the constant controlling the overall strength of the pressure gradient, ω is the angular speed of the oscillating component, $h$ is the channel half-width, and $u_{max}$ is the centerline velocity of the steady flow component. The same parameters were selected as by Charonko et al. (2010) with $\omega = 2\pi$ rad/s (period T = 1 s), $u_{max} = 1\ m/s$, and $\gamma = 25.13$. A flow domain with $h$=4 mm and a length of 20 mm was employed as shown in Fig. 3(a). The pressure along the inflow boundary was set to be 0 Pa, which was employed as Dirichlet BC for solving the PPE and as the reference points for the least-squares based methods. The fluid properties were given as $\rho = 1000\ kg/m^3$ and $\mu = 1 \times 10^{-3}$ Pa·s. The velocity fields were generated on a uniform Cartesian grid with a grid size of 0.1 mm, yielding 101×41 grid points. For each test case, 1000 velocity fields were generated with a span of 50 cycles at a sampling rate of 20 Hz. The waveforms of streamwise centerline velocity and pressure gradient are shown in Fig. 3(b) for one cycle. The streamwise velocity profiles at 4 phases are shown in Fig. 3(c). The characteristic velocity and pressure are defined as $u_0 = u_{max}$ and $p_0 = \frac{1}{2}\rho u_0^2$, respectively.

To test the robustness of pressure reconstruction methods, Gaussian noise with different levels of spatial auto-correlation was added to the true velocity fields. Spatially correlated errors have been reported by Sciacchitano and Wieneke (2016) for PIV fields and are anticipated in gridded velocity fields interpolated from PTV measurements since the error of each single particle track can affect the velocity values on multiple grid points. Three levels of spatial correlation were considered, namely uncorrelated (UC), weakly correlated (WC), and strongly correlated (SC). The spatial auto-correlation coefficient was specified as

$$\rho_{u1,u2} = \exp\left(-s\left(\frac{r_{1,2}}{\Delta x}\right)^2\right), \tag{23}$$

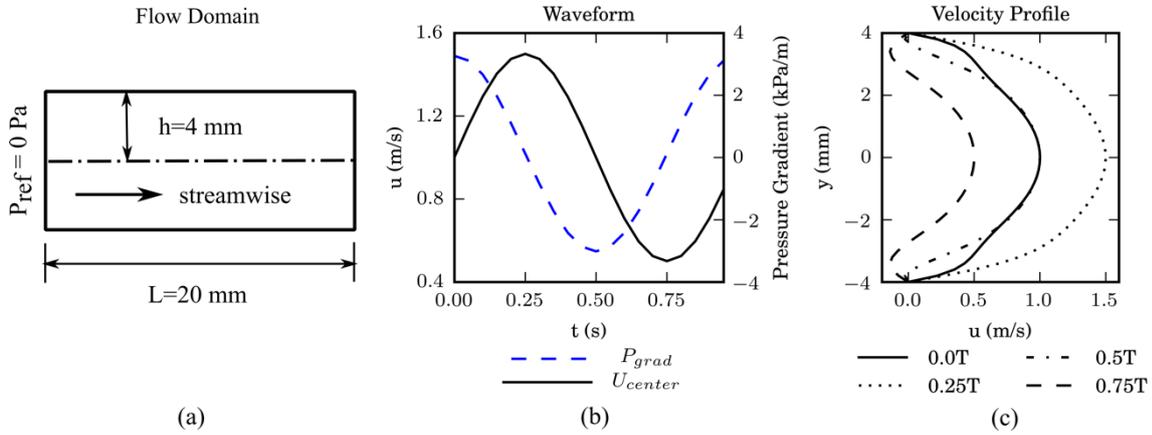

**Fig. 3 (a) The domain arrangement of the 2D pulsatile flow. (b) The streamwise centerline velocity and pressure gradient waveforms within a cycle. (c) The streamwise velocity profile at 4 temporal phases.**

where $\rho_{u1,u2}$ is the auto-correlation coefficient between velocity errors at two points denoted as 1 and 2, $r_{1,2}$ is the spatial distance between the two pints, $\Delta x$ is the grid size, and $s$ is a positive constant controlling the strength of the correlation. $s$ is zero for UC errors, and was set at 0.22 and 0.88 for WC and SC errors, respectively. The SC $\rho_{u1,u2}$ was similar to the results of PIV measurements with 32 × 32 pixels interrogation window and 75% overlap (Sciacchitano and Wieneke 2016). The covariance was then calculated based on the auto-correlation coefficient as

$$Cov_{u1,u2} = \rho_{u1,u2}\sigma_1\sigma_2, \qquad (24)$$

where $Cov_{u1,u2}$ is the covariance between velocity errors at points 1 and 2 whose variances are $\sigma_1$ and $\sigma_2$. The velocity error variance was defined as a fraction of the true velocity magnitude at each point as

$$\sigma = \alpha|\boldsymbol{u_t}|, \qquad (25)$$

where $\alpha$ controls the level of imposed errors. Assuming the errors of one velocity component are uncorrelated with the other components, and the errors at different frames are uncorrelated, the covariance matrix $\Sigma_u$ was constructed based on the specified variances and covariances, then the spatially correlated velocity errors were generated by multiplying the Cholesky decomposition of $\Sigma_u$ to a vector containing uncorrelated, unbiased, and Gaussian distributed random noise with unity standard deviation (Azijli and Dwight 2015). To test the pressure reconstruction methods for a wide range of error levels, 11 test cases were created for each correlation level with $\alpha$ varying from 1% to 50%, resulting 33 test cases in total.

### 2.3 Laminar pipe flow measurement

The laminar flow in a circular pipe was measured using a volumetric PTV experiment and was employed to validate the GLS pressure reconstruction method. The schematic illustration of the experimental setup is shown in Fig. 4, and more details about the experiment can be found in the work by Bhattacharya and Vlachos (2019). The flow was driven using a gear pump with a steady flow rate Q of 0.17 L/min. The flow rate upstream and downstream of the pipe was measured using an ultrasonic flowmeter and the average flow rate was used to determine the true velocity profile. A clear FEP tube of diameter ($R_{pipe}$) 0.25 inch was used for the experiment. The working fluid inside the pipe was distilled water-urea (90:10) solution with a density of 1015 kg/m³ and dynamic viscosity of 0.915 mPas. The pipe flow has a Reynolds number of 630 and was in the laminar flow regime. The pipe was also immersed within the water-glycerol solution such that it is refractive index matched. The measurement volume was 9×6.5×6.5 mm³ and was illuminated by a continuum Terra-PIV laser with appropriate optical setup. The time-resolved measurements were taken at 6 kHz with 4 cameras, and the image size was 640×624 pixels. 24-micron fluorescent particles were used with a particle Stokes number of 0.0005. The particle images were processed using in-house camera calibration, particle reconstruction and tracking code. A polynomial mapping function (Soloff et al. 1997) was used to establish a relation between image coordinates and physical coordinates. Three iterations of volumetric self-calibration (Wieneke 2008) were done to eliminate any disparity between the measurement volume and calibration target location or alignment. The 3D triangulation (Maas et al. 1993) was used to reconstruct the particle positions in physical coordinate system and subsequently the 3D particle locations were tracked using a "nearest-neighbor" pairwise tracking algorithm. A total of 499 velocity frames were obtained from 500 snapshots of particle images. The velocity values at particle locations were interpolated to a uniform Cartesian grid using "FlowFit" (Gesemann et al. 2016). The grid resolution was 0.385 mm, and the ratio between the number of tracked particles with the number of grid points in the flow domain was 0.2. Some measurements at the pipe wall were trimmed to avoid the significant errors due to lack of particles in those regions.

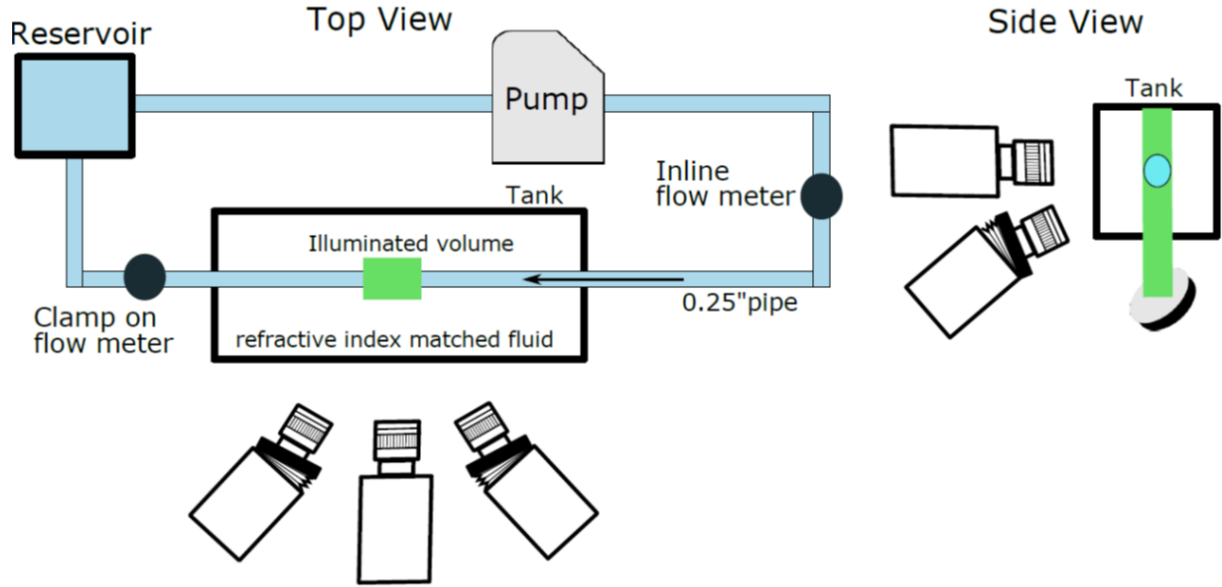

**Fig. 4 Schematic of laminar pipe flow set up showing the flow loop and camera arrangement.**

The proposed GLS method and the PPE method were applied to the gridded velocity fields. A zero reference pressure was assigned at the center point of inflow plane, while Neumann BC was given at the rest of the boundaries with the pressure gradients calculated from the velocity data. The velocity standard deviation (STD) between all the 499 frames were calculated at each spatial point, which was then utilized to generate the covariance matrix $\Sigma_{u,STD}$ for GLS reconstruction. In addition, another set of covariance matrix $\Sigma_{u,UNC}$ was generated from $\hat{\sigma}_u$ estimated using the velocity-divergence based algorithm introduced in Sect. 2.1.5. The covariances were assumed to be zero in both $\Sigma_{u,STD}$ and $\Sigma_{u,UNC}$. The GLS reconstructions with $\Sigma_{u,STD}$ and $\Sigma_{u,UNC}$ are denoted as GLS STD and GLS UNC, respectively.

The analytical solution of the laminar pipe flow is:

$$U = -\frac{1}{4\mu}\frac{dP}{dx}\left(R_{pipe}^2 - R^2\right), \text{ with}$$

$$\frac{dP}{dx} = \frac{8\mu Q}{\pi R_{pipe}^4}, \tag{26}$$

where x is the streamwise direction, R is the radial distance, and U is the streamwise velocity. The centerline velocity magnitude $U_{centerline}$ was employed as the characteristic velocity, and the characteristic pressure $p_0$ was $\frac{1}{2}\rho U_{centerline}^2$.

## 3 Results

**3.1 Synthetic flow fields**

*3.1.1 Pressure gradient uncertainty estimation*

For the 2D pulsatile flow fields, the relative velocity error magnitudes ($|\epsilon_u|$) were calculated as $\sqrt{\epsilon_u^2 + \epsilon_v^2}/u_0$. The overall velocity error level for each test case was represented as the median $|\epsilon_u|$ from all the points in space and time. The pressure gradient fields were calculated as introduced in Sect. 2.1.1, and the relative pressure gradient errors $\epsilon_{\nabla p}$ were evaluated as the deviations from the analytical solutions, then normalized by $p_0/\Delta x$. To validate the linear-transformation based uncertainty propagation algorithm for both instantaneous and local prediction, the root mean square (RMS) value of the estimated uncertainty distributions were compared with the RMS of the true error distributions in time and space since the RMS error should match the RMS uncertainty for a successful prediction (Sciacchitano et al. 2015). In Fig. 5(a), the temporal variations of the RMS values of the estimated relative streamwise pressure gradient uncertainty ($\hat{\sigma}_{pgradx}$) and the relative streamwise pressure gradient errors ($|\epsilon_{pgradx}|$) were compared for the test case with 9.6% SC velocity errors ($\alpha$=15%) at two spatial locations (along the centerline and at R=3mm), and the

RMS values of $\hat{\sigma}_{pgradx}$ were consistent with those of $|\epsilon_{pgradx}|$ for both spatial locations at all time points. In Fig. 5 (b), the spanwise distributions of the RMS values of $|\epsilon_{pgradx}|$ and $\hat{\sigma}_{pgradx}$ were compared at two phases (t/T=0.25 and 0.75). Greater $|\epsilon_{pgradx}|$ and $\hat{\sigma}_{pgradx}$ were found near the centerline of the flow field for both phases, and the RMS values of $\hat{\sigma}_{pgradx}$ were consistent with those of $|\epsilon_{pgradx}|$. Overall, the comparisons suggested that the proposed algorithm was capable of predicting the instantaneous and local uncertainty of the calculated pressure gradients.

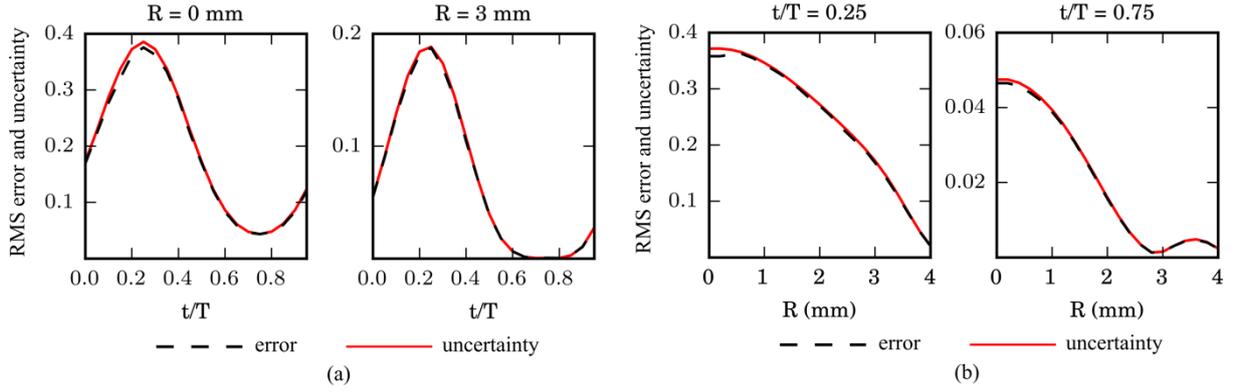

**Fig. 5 For the case with 9.6% SC velocity errors, (a) The RMS of streamwise relative pressure gradient errors $\epsilon_{pgradx}$ and uncertainties $\hat{\sigma}_{pgradx}$ at all temporal phases for the grid points along centerline (R=0 mm) and at R=3 mm. (b) The spanwise distributions of RMS errors and uncertainties at t/T=0.25 and at t/T=0.75.**

The estimated auto-correlation coefficients $\hat{\rho}_{pgradx}$ were validated by comparing to the statistically quantified coefficients $\rho_{pgradx}$ based on the true errors. For each test case and each time phase, the $\rho_{pgradx}$ between all pairs of spatial points were quantified using the errors across all the frames as

$$\rho_{pgradx} = \frac{Cov(\epsilon_{pgradx,1}, \epsilon_{pgradx,2})}{\sigma_{pgradx,1}\sigma_{pgradx,2}} \tag{27}$$

where $Cov(\epsilon_{pgradx,1}, \epsilon_{pgradx,2})$ is the covariance between pressure gradient errors at points 1 and 2, while $\sigma_{pgradx,1}$ and $\sigma_{pgradx,2}$ are the STDs of pressure gradient errors at the two points, respectively. The median absolute $\rho_{pgradx}$ was determined to illustrate the auto-correlation strength for each spatial separation r. As a demonstration, the median absolute $\rho_{pgradx}$ as a function of $r/\Delta x$ is shown in Fig. 6(a), for the test cases with 9.6% velocity errors. The estimated coefficients on the right quadrant were consistent with the quantified coefficients on the left quadrant. In general, the auto-correlation of $\epsilon_{pgradx}$ was stronger for smaller r as well as for the test case with stronger correlated $\epsilon_u$. At $r = \Delta x$, the median absolute $\rho_{pgradx}$ was 0.07 for UC case, while the values were 0.32 and 0.71 for WC and SC cases, respectively. The corresponding $\hat{\rho}_{pgradx}$ values were 0.03 for UC, 0.36 for WC, and 0.72 for SC. To validate the local and instantaneous prediction of auto-correlation coefficients, the $\rho_{pgradx}$ and $\hat{\rho}_{pgradx}$ values between the center point (x=10 mm and y=0 mm) and its neighboring points are presented in Fig. 6 (b) and (c), respectively, for the test case with 9.6% SC velocity errors at phase t/T=0.25. The estimated values were also in good agreement with quantified results. The $\rho_{pgradx}$ decreased monotonically from the center point along both spanwise and streamwise directions, and the decreasing rate was greater along streamwise direction than the spanwise direction with negative coefficients near the edges. The $\rho_{pgradx}$ distributions were also investigated for other regions as well as other phases of the flow and were found to be dependent on the local and instantaneous flow conditions. The proposed uncertainty estimation algorithm was able to estimate the $\rho_{pgradx}$ accurately for all the investigated locations and phases.

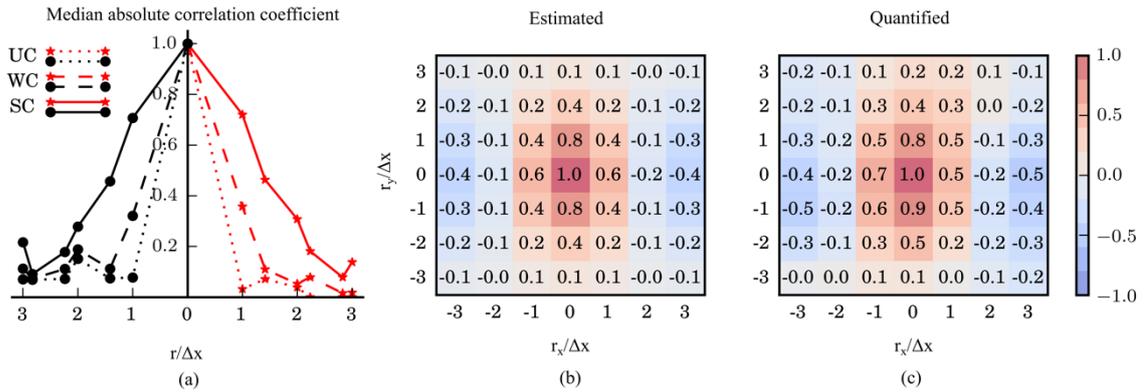

**Fig. 6 (a) The median absolute auto-correlation coefficients of $\rho_{pgradx}$ as a function of normalized spatial separation $r/\Delta x$. The statistically quantified $\rho_{pgradx}$ from $\epsilon_{pgradx}$ (left quadrant) is compared with the estimated $\hat{\rho}_{pgradx}$ using proposed linear-transformation based algorithm (right quadrant). The distributions of $\hat{\rho}_{pgradx}$ (b) and $\rho_{pgradx}$ (c) around the center point at t/T=0.25.**

### 3.1.2 Pressure reconstruction

The instantaneous pressure fields were reconstructed using the methods introduced in Section 2.1. From each test case, the pressure errors were calculated as the deviations from analytical solution at all the points in space and time, then normalized by $p_0$. Three pressure error metrics were employed to evaluate the performances of the methods. The median absolute pressure error was used to represent the overall pressure error level, while the 15.75[th] and 84.25[th] percentiles were defined as the lower-bound (LB) and upper-bound (UB), respectively. The pressure error metrics as functions of the velocity error levels are compared in Fig. 7(a) between PPE and GLS, and in Fig. 7(b) between least-squares based methods. In addition, the pressure error level was also determined based on the bias error and the random error separately. The bias error at temporal phase $\phi$ and spatial location $r$ was quantified as $\epsilon_{bias}^{\phi,r} = \frac{1}{N_{cycle}} \sum_{k=1}^{N_{cycle}} \epsilon_p^{k,\phi,r}$, where $N_{cycle} = 50$ is the total number of cycles, and $\epsilon_p^{k,\phi,r}$ is the pressure error at cycle $k$, temporal phase $\phi$, and spatial location $r$. The random error was then evaluated by subtracting the bias error from the total error as $\epsilon_{random}^{k,\phi,r} = \epsilon_p^{k,\phi,r} - \epsilon_{bias}^{\phi,r}$. As shown in Fig. 7, with $\alpha$ varied from 1 % to 50 %, the velocity error level increased from 0.64% to 32.1%. The GLS method was more robust to velocity errors compared with the other methods. For the case with 32.1% UC velocity errors, the total pressure error was 20.3% by PPE and only 5.8% by GLS as suggested in Fig. 7(a). For the same case, WLS and OLS yielded 6.9% and 20.4% pressure errors, respectively, as suggested in Fig. 7(b). The error bounds of GLS were also lower than those by PPE, OLS, and WLS. The improvement by GLS was more significant for cases with greater velocity errors. Compared to PPE, the GLS method reduced the total pressure error level by 50% (2.4% vs 3.6%) with 9.6% UC velocity errors, and by 250% (5.8% vs 20.3%) with 32.1% UC velocity errors.

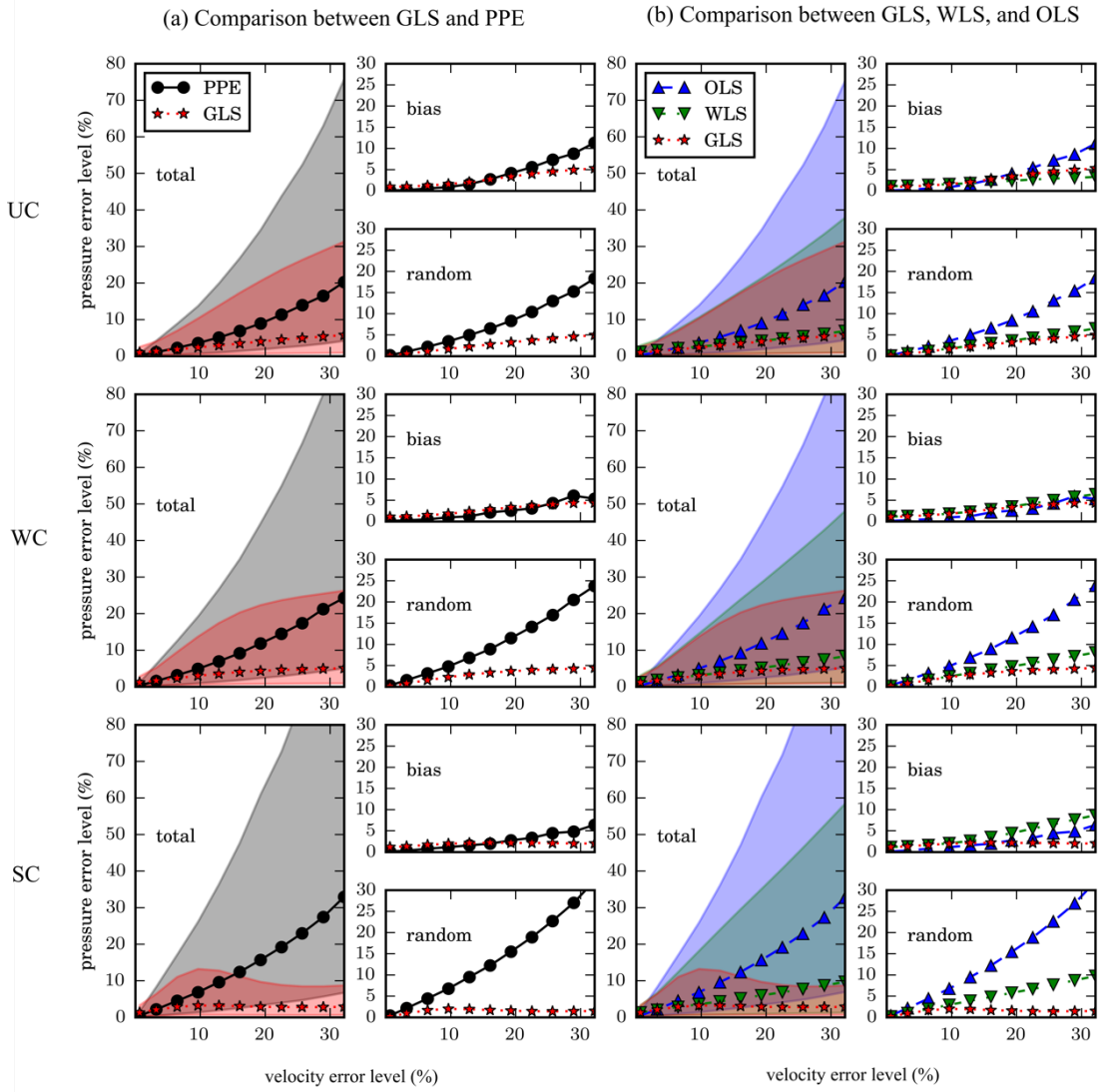

**Fig. 7 Comparisons of the pressure reconstruction methods for a wide range of velocity error levels and three correlation levels. (a) Comparison between GLS and PPE. (b) Comparison between GLS, WLS, and OLS. From top to bottom, the correlation levels are uncorrelated (UC), weakly correlated (WC), and strongly correlated (SC).**

The spatial correlation of velocity errors amplified the pressure errors by PPE, OLS, and WLS, as suggested in Fig. 7. With 32.1% velocity errors, the pressure error levels by PPE as 20.3% for UC, 24.3% for WC, and 32.9% for SC. In contrast, GLS yielded lower pressure errors for cases with stronger correlated velocity errors. With 32.1% velocity errors, the pressure error levels were 5.8 for UC, 5.1% for WC, and only 2.9% for SC. At around 10% SC velocity level, the pressure error level by GLS plateaued, and the increase of velocity error level no longer amplified the resulted pressure error level. As a consequence, the improvement by GLS was more significant for cases with spatial correlated velocity errors.

The WLS had similar performances as GLS for cases with UC velocity errors, but not for WC or SC cases as shown in Fig. 7(b).

However, the GLS method yielded greater pressure errors for cases with minimal velocity errors compared to PPE and OLS. As suggested in Fig. 7, the total pressure error level by GLS was 0.96% with 0.64% UC velocity errors, while it was 0.24% and 0.27% by PPE and OLS, respectively. GLS created more bias errors for cases with low velocity errors compared to OLS and PPE, while the random errors were consistently reduced by GLS across all the cases. Since the random errors were more significant than bias errors in most cases, the GLS reduced the overall pressure error levels. For the cases with higher velocity error levels (greater than 20%), GLS reduced both bias and random errors, therefore improved the pressure accuracy significantly.

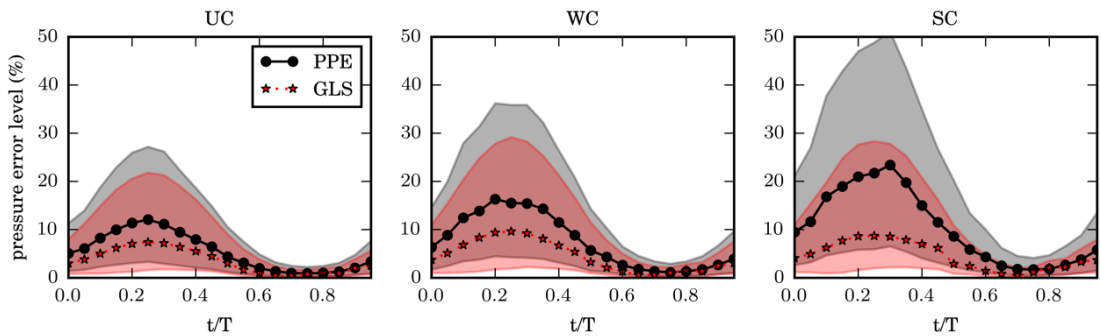

**Fig. 8 Comparison between GLS and PPE at all time phases for the cases with 9.6% velocity error level.**

The performance of GLS was investigated at different time phases of the pulsatile flow. Fig. 8 compares the pressure error levels and error bounds between GLS and PPE at each phase for the test cases with 9.6% velocity errors. The pressure errors were greater for the phases with stronger flow. With SC velocity errors, the pressure error level by GLS was 3.2% for the whole cycle, while it was 8.7% at t/T=0.25 (maximum flow rate) and 1.2% at t/T=0.75 (minimum flow rate). For phases with low flow rates (t/T between 0.6 to 0.8), both GLS and PPE yielded accurate pressure fields with error levels less than 5%. However, GLS improved the pressure accuracy significantly for the phases with greater flow rates. At t/T=0.25 with SC velocity errors, the pressure error level by PPE was 21.7% which was 149% greater than GLS.

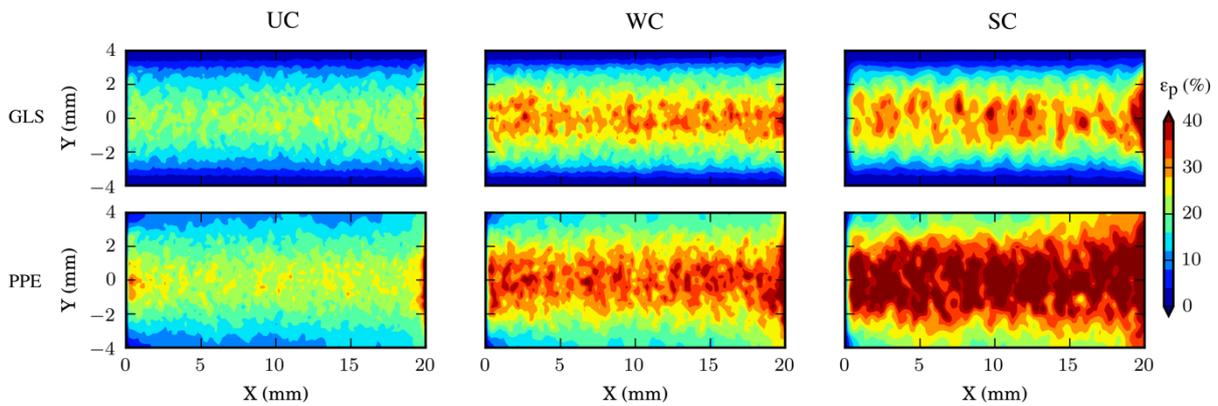

**Fig. 9 The spatial distributions of pressure RMS errors by GLS (top row) and PPE (bottom row) for the test cases with 9.6% velocity error level.**

The spatial distributions of relative RMS pressure errors are compared in Fig. 9 between GLS and PPE at phase t/T=0.25 for the cases with 9.6% velocity errors. The RMS pressure errors were greater near the centerline of the flow field due to the greater pressure gradient errors as suggested in Fig. 5(b). Compared to PPE, the GLS constrained the high pressure errors within the centerline region and dramatically reduced the pressure errors in near-wall regions,. In addition, the improvement by GLS was more significant for test cases with spatial correlated velocity errors. With SC errors, the GLS reduced the pressure errors significantly for both centerline region and the near-wall regions.

The GLS pressure reconstruction was also performed for WC and SC cases with the assumption of zero velocity error covariances (GLS 0Cov). Compared to GLS, the GLS 0Cov underestimated the covariances of $\epsilon_{\nabla p}$ due to the ignorance of the auto-correlation of velocity errors, and the captured $\epsilon_{\nabla p}$ covariances were caused by the numerical differentiations during pressure gradient calculation. The pressure error levels as functions of velocity error levels were compared between PPE, GLS, and the GLS 0Cov in Fig. 10. As the velocity error level increased from 0.64% to 32.1%, the pressure error level by GLS 0Cov changed from 1.0 to 7.1 % for WC cases and from 1.0 to 8.8 % for SC cases. By neglecting the velocity error covariances, the GLS 0Cov had a slightly worse performance than GLS for WC cases, while the performance deficit became more dramatic for SC cases due to the significance of velocity error auto-correlations. With 16.1% velocity errors, the pressure error level by GLS 0Cov was 8% larger than GLS for the WC case and 70% larger for the SC case. Compared to the baseline method PPE, GLS 0Cov still reduced the pressure errors dramatically for most cases. With 16.1% velocity errors, the pressure error reduction by GLS 0Cov was 110% for the WC case and 138% for the SC case.

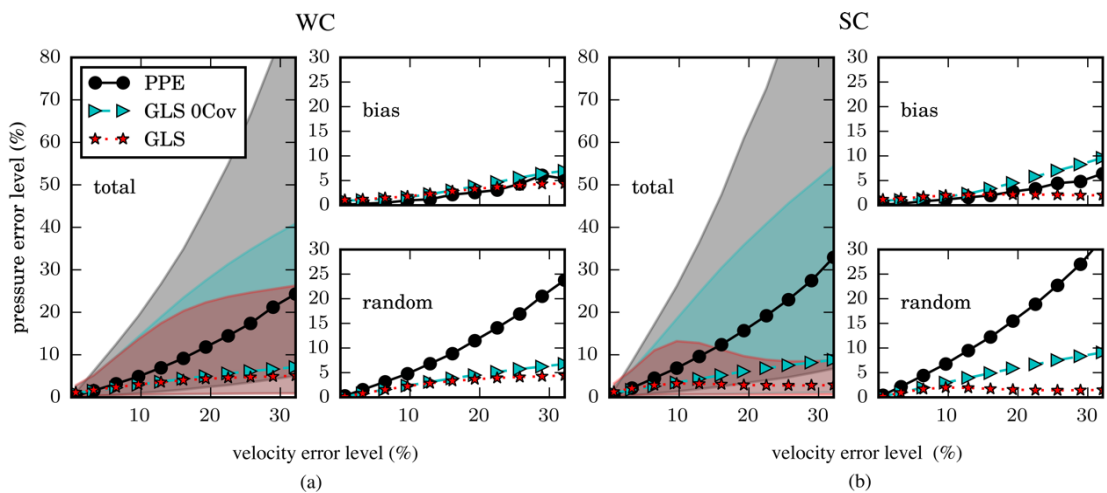

**Fig. 10 Comparisons between pressure reconstruction using PPE, GLS with full velocity error covariances, and GLS with zero velocity error covariances for the test cases with WC velocity errors (a) and SC velocity errors (b).**

### 3.2 Laminar pipe flow

For the experimental validation case of laminar pipe flow, the parabolic streamwise (along X direction) velocity profile is shown in Fig. 11(a) as a function of Y and Z using the averaged values across all the frames and Y-Z slices. The velocity errors were calculated as the deviations between the gridded velocity data and the analytical solution given by Eqn. 26 then normalized by the characteristic velocity $U_{centerline}$. To assess the accuracy of the measurement, the histograms of the relative velocity error magnitudes ($|\epsilon_u|$), velocity STD magnitudes ($|\sigma_u|$), and velocity uncertainty magnitudes ($|\hat{\sigma}_u|$) estimated using the divergence-based algorithm introduced in 2.1.5 are shown in Fig. 11(b). The RMS values was 12.0% for $|\epsilon_u|$, 10.9% for $|\sigma_u|$, and 5.9% for $|\hat{\sigma}_u|$, as suggested by the vertical lines. The spatial distributions of $|\sigma_u|$ and $|\hat{\sigma}_u|$ as functions of radial location $R = \sqrt{Y^2 + Z^2}$ and streamwise location X were shown in Fig. 11(c), and greater $|\sigma_u|$ was found near the wall of the pipe. In general, the $|\hat{\sigma}_u|$ had a 50% underprediction and less spatial variation. The relative errors of the calculated pressure gradients $\epsilon_{\nabla p}$ were determined as the deviations from the analytical solution then normalized by $p_0/\Delta x$. Two sets of pressure gradient uncertainties were estimated using the proposed linear-transformation based algorithm from $\sigma_u$ and $\hat{\sigma}_u$, and are denoted as $\hat{\sigma}_{\nabla p}(\sigma_u)$ and $\hat{\sigma}_{\nabla p}(\hat{\sigma}_u)$, respectively. The histograms of the relative errors and uncertainties are shown in Fig. 11(d). The RMS values were 198% for $|\epsilon_{\nabla p}|$, 200% for $|\hat{\sigma}_{\nabla p}(\sigma_u)|$, and 109% for $|\hat{\sigma}_{\nabla p}(\hat{\sigma}_u)|$ as suggested by the vertical lines. The spatial distributions of the STD of pressure gradients ($\sigma_{\nabla p}$) and the uncertainties are shown in Fig. 11(e) as functions of R and X. The $\hat{\sigma}_{\nabla p}(\sigma_u)$ was consistent with the $\sigma_{\nabla p}$ in terms of the histograms and the spatial distributions, while $\hat{\sigma}_{\nabla p}(\hat{\sigma}_u)$ yielded an underprediction and less spatial variation since it was based on $\hat{\sigma}_u$.

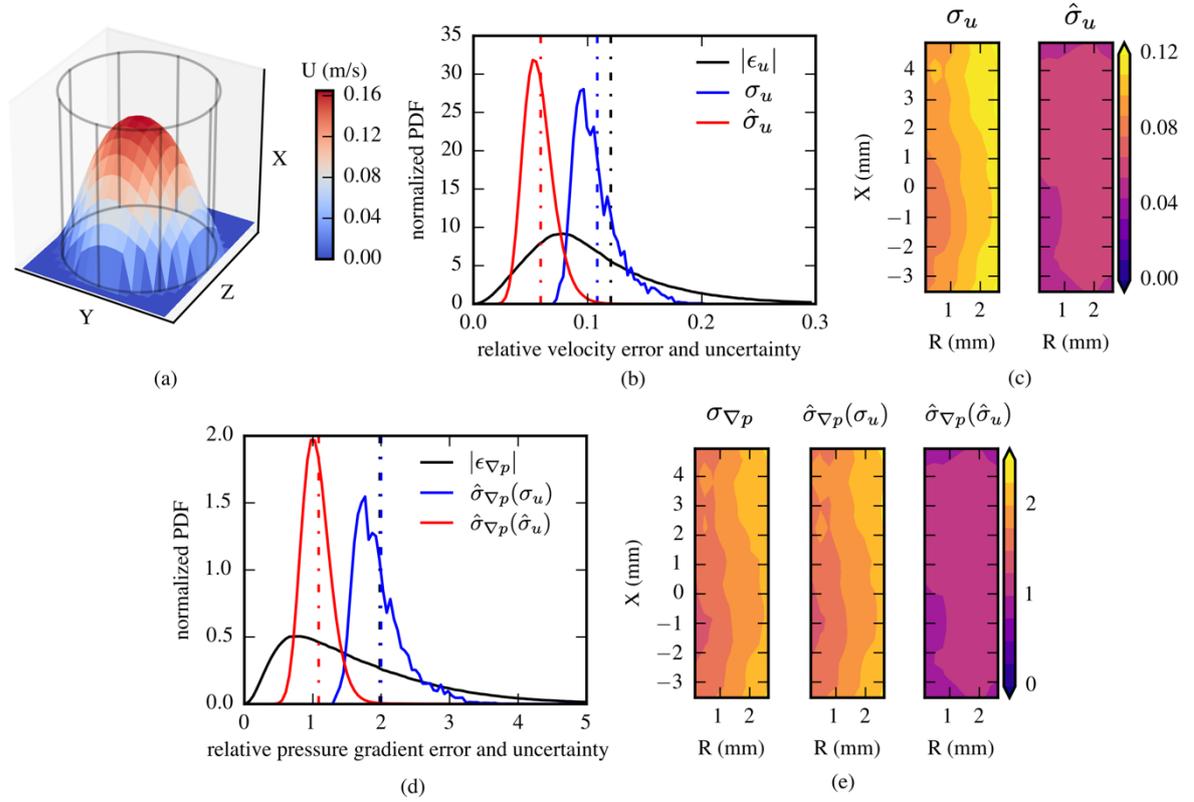

Fig. 11 (a) The average streamwise velocity profile as a function of Y and Z. (b) The histograms of the magnitudes of velocity errors ($|\epsilon_u|$), standard deviations ($\sigma_u$) and estimated uncertainties ($\hat{\sigma}_u$). The vertical lines represent the RMS values of the distributions. (c) The spatial distributions of $\sigma_u$ and $\hat{\sigma}_u$. (d) The histograms of the magnitudes of pressure gradient errors ($|\epsilon_{\nabla p}|$), the uncertainties estimated from $\sigma_u$ ($\hat{\sigma}_{\nabla p}(\sigma_u)$), and the uncertainties estimated from $\hat{\sigma}_u$ ($\hat{\sigma}_{\nabla p}(\hat{\sigma}_u)$). The vertical lines represent the RMS values of the distributions. (e) The spatial distributions of the magnitudes of pressure gradient STD ($\sigma_{\nabla p}$) and the estimated uncertainties $\hat{\sigma}_{\nabla p}(\sigma_u)$ and $\hat{\sigma}_{\nabla p}(\hat{\sigma}_u)$.

The errors of the reconstructed pressure fields were quantified as the deviations from the analytical solution then normalized by the characteristic pressure $p_0$. The pressure STDs over the 499 velocity frames were also calculated to assess the precision of the pressure results. The histograms of the absolute pressure errors ($|\epsilon_p|$) and pressure STDs ($\sigma_p$) were compared in Fig. 12 (a) and (b), respectively. The median of $|\epsilon_p|$ was 219% for PPE, 112% for GLS STD, and 135% for GLS UNC, as suggested by the vertical dashed lines in Fig. 12(a). Compared to PPE, the error reduction was 96% by GLS STD and 62% by GLS UNC in terms of median $|\epsilon_p|$. The median of $\sigma_p$ was 308% for PPE, 165% for GLS STD, and 200% for GLS UNC, which suggested a precision improvement of 87% by GLS STD and 54% by GLS UNC. The spatial distributions (as functions of R and X) of the RMS $\epsilon_p$ were compared in Fig. 12(c) between PPE and GLS reconstructions. The pressure errors were lower near the center of inflow plane where the reference pressure was assigned, and increased towards the outflow plane and pipe walls. The GLS method mitigated the error propagation across the field and therefore effectively improved the pressure accuracy.

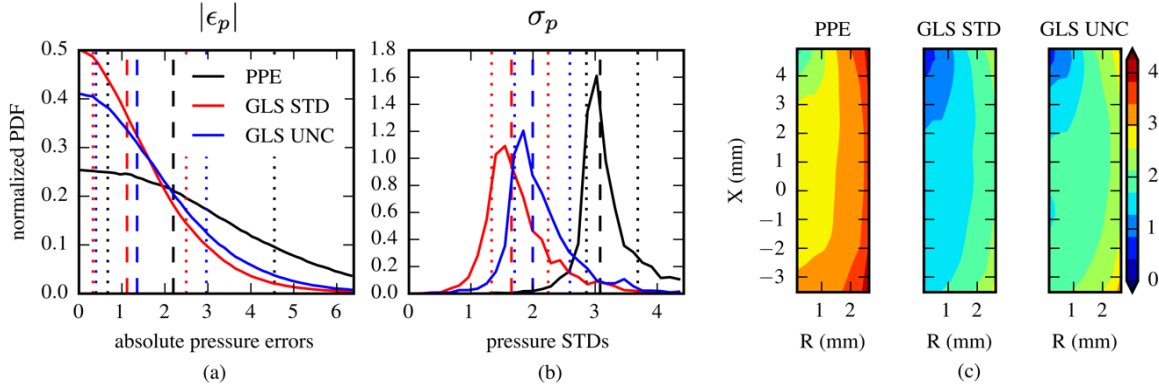

**Fig. 12 Histograms of the absolute pressure errors $|\epsilon_p|$ (a) and pressure STDs $\sigma_p$ (b) by PPE and the GLS reconstructions with different velocity uncertainty sources (GLS STD with $\sigma_u$ and GLS UNC with $\hat{\sigma}_u$). The vertical dashed lines represent the medians, while the vertical dotted lines indicate the LBs and UBs of the distributions. (c) The spatial distributions of pressure RMS errors as functions of R and X.**

## 4  Discussion

In this study we introduced a novel pressure reconstruction method using uncertainty-based generalized least-squares. By propagating the velocity uncertainty with the proposed linear-transformation based algorithm, the pressure gradient uncertainty are estimated in the form of covariance matrices of the heteroscedastic and correlated pressure gradient errors. The pressure integration was formulated as solving an overdetermined linear system involving pressure gradient fields and boundary conditions. The GLS estimator is the best unbiased linear estimator which provides the unbiased estimation of the governing Eqn. 10 with the lowest variance. The error mitigation by GLS was the result of incorporating the pressure gradient uncertainty into the pressure integration. The OLS method, which solves the same linear system (Eqn. 15) as GLS without utilizing the uncertainty, yielded a similar performance as PPE for all the synthetic flow cases. This suggested that the change of formulation alone (the overdetermined linear system compared to the PPE) does not affect the robustness of pressure reconstruction, which was consistent with the statement by Wang et al. (2017) that PPE and OLS share the same theoretical foundation. The comparisons between OLS, WLS, and GLS in Fig. 7(b) indicated that both the variance and covariance of the pressure gradient errors contributed to the improvement of pressure accuracy. The WLS method, which only employs the variances, had a better performance than OLS. By considering both variances and covariances, the GLS method further improved the pressure accuracy compared to WLS. With stronger correlated velocity errors, the covariances of pressure gradient errors became more significant as suggested by Fig. 6(a), thus the GLS was more effective as it utilized the covariances for pressure reconstruction as suggested in Fig, 7.

The proposed linear-transformation based algorithm was capable of predicting the variances and covariances of the pressure gradient errors locally and instantaneously, as suggested in Figs. 5 and 6. Compared with the approaches introduced by Azijli et al. (2016) which performed exact uncertainty propagation assuming Gaussian-distributed velocity errors or carried out Monte Carlo simulations, the proposed algorithm does not assume any particular form of the velocity error distribution and is more computationally efficient due to the linearization of the error propagation. The formulation of Eqn. 7 and the analyses in Sect. 3.1.1 also suggested that the pressure gradient uncertainty is dependent on many factors including the velocity uncertainty, the discretization scheme, the spatiotemporal resolution, the local velocity profile, etc. As a consequence, the variances and covariances of the pressure gradient errors usually differ from those of velocity errors. One example is the auto-correlation coefficients $\rho_{pgradx}$ around the center point in Fig. 6 (b-c). Although the $\rho_{u1,u2}$ (as defined by Eqn. 23) varied isotropically with nonnegative values, the variation of $\rho_{pgradx}$ was anisotropic with both positive and negative correlations. In addition, the error magnitudes of the calculated pressure gradient can be significantly different from those of the measured velocity. For the laminar pipe flow, the RMS error amplification from velocity to pressure gradient was 16.5 as suggested in Figs. 11(b) and (d). This dramatic error amplification was due to the small time separation $\Delta t$ (1/6000 s) of the flow measurement. As discussed by van Oudheusden (2013), decreasing $\Delta t$ reduces the truncation error of finite difference, but increases the effect of velocity error on pressure gradient calculation with Eulerian approach. For steady flows, using a large $\Delta t$ or assuming zero $\frac{\partial u}{\partial t}$ would effectively reduce the error amplification. However, the pressure reconstruction was still performed with the highest resolved temporal frequency in the present study so that the analysis and conclusions would be applicable for unsteady flows which are more commonly seen in real-world applications.

The performance of GLS pressure reconstruction was affected by the reliability of the provided velocity uncertainty. Since the a-posterior method to estimate velocity uncertainty from volumetric PTV has not been established, the uncertainty of the laminar pipe flow data was obtained from the velocity fields with two approaches in this study. The first approach took the STD over all the velocity frames ($\sigma_u$) and was a reliable estimation for the steady flow. The second approach estimated $\hat{\epsilon}_u$ from the spurious velocity-divergence of the incompressible flow then estimated the $\hat{\sigma}_u$ as the WSTD of $\hat{\epsilon}_u$. As the $\hat{\epsilon}_u$ was calculated from Eqn. 19 in a least-squares sense, an underestimation of $\hat{\sigma}_u$ was caused as suggested in Fig. 11(b). With the less reliable $\hat{\sigma}_u$, the GLS UNC had a slightly worse performance than GLS STD, while both GLS UNC and GLS STD still effectively improved the pressure accuracy compared to PPE.

The performance of GLS was also affected by the completeness of the provided velocity uncertainty. To ensure an accurate prediction of the pressure gradient uncertainty, both the variances and covariances of the velocity errors are required. However, there has not been any established method to estimate the covariances of velocity errors for PIV/PTV. A more practical condition was considered in this study by performing the GLS with zero velocity covariances (GLS 0Cov) on the synthetic flow fields with correlated velocity errors (nonzero covariances). As suggested in Fig. 10, the GLS 0Cov had slightly worse performances than GLS, while still effectively improved the pressure accuracy compared to PPE. For the experimental laminar pipe flow, the velocity error covariances were also unavailable and assumed to be zero, and the GLS pressure reconstructions yielded more accurate pressure estimations than PPE as suggested in Fig. 12. These analyses indicated the practical benefits of using GLS for pressure reconstruction even without the velocity error covariances.

There are several limitations of the uncertainty-based GLS pressure reconstruction method. First, the GLS method requires the velocity uncertainty to be estimated instantaneously and locally. For planar/stereo PIV measurements, there are several methods to estimate the velocity uncertainty (Timmins et al. 2012; Charonko and Vlachos 2013; Xue et al. 2015; Bhattacharya et al. 2017, 2018). However, these methods remain untested for volumetric PIV, and only one recent development has covered the a-posterior uncertainty quantification for volumetric PTV (Bhattacharya and Vlachos 2019). Moreover, none of these methods can provide the covariances of velocity errors. The performance of GLS with the velocity uncertainty estimated using the existing uncertainty quantification methods can be explored in future work. Second, the GLS method has greater computational cost than the other methods employed in the present study. As described in Section 2.1.4, the GLS reconstruction needs to solve the augmented linear system to avoid operations with the dense matrix $\Sigma_b^{-1}$. The augmented systems are approximately 3 times and 4 times as large as the linear systems of PPE for planar data and volumetric data, respectively. With the sparse solver or the iterative methods whose calculation complexity is in $\mathcal{O}(n_{LS}^2)$, where $n_{LS}$ is the size of the linear system, the computational cost of GLS is approximately 9 times as much as solving the PPE for planar data and 16 times for volumetric data. Therefore, the GLS method is better suited with planar fields or smaller volumetric fields (in terms of the number of grid points each frame) such as the laminar pipe flow data which contains 3772 grid points within the flow field. For larger volumetric flow fields, the WLS method is preferred which requires similar computational cost as PPE. The WLS reconstruction was demonstrated with synthetic flow fields and showed significant improvement compared to PPE as suggested in Fig. 7(b). The WLS has also been employed to estimate the instantaneous pressure fields from velocity fields measured using PTV in patient-specific aneurysm models (Zhang et al. 2019). In addition, the proposed GLS pressure reconstruction can only be applied to incompressible flow fields since the formulations of pressure gradient calculation and uncertainty propagation are only valid for incompressible flows. Also, the current framework of GLS is only applicable to gridded velocity data and the Eulerian approach for pressure gradient calculation. To ensure the accuracy of numerical differentiations, the GLS method requires spatiotemporal resolved velocity measurements.

## 5 Conclusions

This study presented an uncertainty-based instantaneous pressure reconstruction method using generalized least-squares. The pressure gradient fields were calculated from the velocity fields measured by PIV/PTV, and a linear-transformation based algorithm was introduced to estimate the local and instantaneous pressure gradient uncertainty-based on the velocity uncertainty. The pressure fields are reconstructed by GLS which utilizes the pressure gradient and its uncertainty. The performance of GLS was tested for synthetic flow fields with a wide range of velocity error levels and both correlated and uncorrelated velocity errors. The GLS method effectively reduced the random errors in the pressure fields compared to the baseline method of solving the PPE. The error mitigation by GLS is due to the utilization of both variances and covariances of the pressure gradient errors. The improvement by GLS was more significant for cases with greater velocity errors, and the total error reduction was as much as 250%. With spatially correlated velocity errors, the GLS was more effective as it utilized the stronger correlation of pressure gradient errors during reconstruction. The laminar pipe flow measured by volumetric PTV was employed to demonstrate the GLS pressure reconstruction, and a 96% error reduction was achieved compared to PPE. Overall, the present study successfully demonstrates the framework of employing uncertainty information to improve the pressure reconstruction from 2D or 3D velocity fields. Further development could include the usage of estimated velocity uncertainty from flow measurements for the GLS pressure reconstruction.